# EPICS: A RETROSPECTIVE ON PORTING iocCore TO MULTIPLE OPERATING SYSTEMS

M.R. Kraimer*, J.B. Anderson*, ANL, Argonne IL 60439, USA
J.O. Hill, LANL, Los Alamos NM 87545, USA
W.E. Norum, University of Saskatchewan, Saskatoon, Canada

*Abstract*

An important component of EPICS (Experimental Physics and Industrial Control System) is iocCore, which is the core software in the IOC (input/output controller) front-end processors. At ICALEPCS 1999 a paper was presented describing plans to port iocCore to multiple operating systems. At that time iocCore only supported vxWorks, but now it also supports RTEMS, Solaris, Linux, and WinNT. This paper describes some key features of how iocCore supports multiple operating systems.

## 1 INTRODUCTION

Originally input/output controller (IOC) meant a VME/VXI-based system, which used the vxWorks operating system. The ICALEPCS99 paper [1] described the plan for removing the VME/VXI and vxWorks dependencies. The basic plan was successfully followed, although many details changed.

The iocCore port involved the following major changes to EPICS Base:

- VME/VXI dependencies – These were solved by unbundling all hardware-specific support, i.e., it is all now built as separate products. This support can still be used on VME/VXI vxWorks systems.
- VxWorks libraries – This was solved by defining operating system independent (OSI) interfaces. Generic or operating-system-specific code implements the interfaces.
- Registry – vxWorks provided a facility that allowed iocCore to locate record/device/driver support during initialization. The registry now provides the same facility.
- Build Environment – The old build environment had separate makefiles for the host and for vxWorks targets while the new build environment uses a single makefile for all targets.
- Shell – iocsh together with the registry provides features previously supplied by the vxWorks shell.
- Interrupt level support – vxWorks made it very easy to interact with interrupt handlers. Parts of the original iocCore required interrupt level support. The new iocCore removes these requirements.

## 2 OVERVIEW OF CHANGES

### 2.1 OSI Interfaces

Table 1 provides a brief summary of the OSI interfaces used by iocCore. The last column shows the different implementations required for the supported platforms.

Table 1: OSI Interfaces

| Name | Replaces | Implementation |
|---|---|---|
| epicsRing | rngLib | Generic |
| epicsTimer | wdLib, osiTimer | Generic |
| epicsAssert | epicsAssert | default, vxWorks |
| epicsEvent | semLib | RTEMS, WIN32, POSIX, vxWorks |
| epicsFindSymbol | symFindByName | Default, vxWorks |
| epicsInterrupt | intLib | RTEMS, default, vxWorks |
| epicsMutex | semLib | RTEMS, WIN32, POSIX, vxWorks |
| epicsThread | taskLib | RTEMS, WIN32, POSIX, vxWorks |
| epicsTime | tickLib, osiTime | RTEMS, WIN32, POSIX, vxWorks |
| osiPoolStatus | memLib | RTEMS, WIN32, default, vxWorks |
| osiProcess | osiProcess | RTEMS, WIN32, POSIX, vxWorks |
| osiSigPipeIgnore | osiSigPipeIgnore | WIN32, default, POSIX, vxWorks |
| osiSock | osiSock | Linux, RTEMS, WIN32, default, solaris, vxWorks |

__________
* Work supported by U.S. Department of Energy, Office of Basic Energy Sciences under Contract No. W-31-109-ENG-38.

## 2.2 Registry

vxWorks provides a function, symFindByName, which is used to dynamically locate global data and functions. This facility is unique to vxWorks and is not easily recreated in other environments. Instead a facility to register and find pointers to functions and structures is provided.

When databases are loaded, all record, device, and driver support is located dynamically. This requires that the support must first be registered. While building an IOC application, a Perl program is run. This program reads the database definition file that will be loaded at runtime by dbLoadDatabase and generates a C function to register the record, device, and driver support. This function is linked with the application and is called before IOC initialization.

## 2.3 Build Environment

Because iocCore is now built for multiple architectures, extensive changes were required for the EPICS build system. Although the new system has more functionality, it is simpler than the old system. In each source directory a single Makefile appears instead of three (Makefile, Makefile.Vx, and Makefile.Host).

## 2.4 Shell

The EPICS IOC shell (iocsh) is a simple command interpreter, which provides a subset of the capabilities of the vxWorks shell. It is used to interpret startup scripts and to execute commands entered at the console terminal. It can execute any command that is registered.

iocCore provides many test commands that are automatically registered.

## 2.5 Interrupt Level Support

The vxWorks intLock/intUnlock routines were an essential part of iocCore. Most operating systems do not allow such tight coupling between interrupt routines and user processes. In iocCore most code was modified so that intLock/intUnlock are no longer required. For code that still needs this capability, epicsInterrupt is provided. For operating systems like vxWorks, in which everything runs in a shared memory, multithreaded kernel environment, an implementation of epicsInterrupt is provided. For other operating systems a default version is provided, which uses a global lock (global means global to the iocCore process).

## 3 STATUS OF PORT

The current production release of EPICS base is EPICS base R3.13.5. EPICS base R3.14.x [2] releases are the releases containing the iocCore port.

## 3.1 Work Completed

The first beta release of 3.14 is now available. It supports the following targets:
- VxWorks 5.4
- RTEMS 4.6 – An open source real-time operating system.
- Solaris – Tested on Solaris 8.
- Linux – Tested on Redhat 6.2 and 7.1.
- Win32 – Tested on winNT, win98 and win2000.

## 3.2 Work Remaining

No new major functionality will be added to the 3.14 EPICS base releases, only bug fixes. With more testing we will be ready for the first regular release of 3.14. At this time we recommend that operational systems start migrating to 3.14.

For existing EPICS sites, one major platform is still missing: HPUX. Work is currently in progress to support HPUX11.

## 3.3 Hardware Support

As mentioned previously, the hardware-specific support that was included with R3.13 is now unbundled. Much of the VME support has been built and tested with R3.14 but only works on vxWorks.

An unbundled version of the sequencer, which was originally developed by William Lupton at KECK and is now supported by Ron Chestnut at SLAC, has been available for some time. This is the version that is supported with R3.14. It has been built and tested on all supported R3.14 platforms.

The unbundled version of GPIB support, which has been available from Benjamin Franksen at BESSY, is the version supported with R3.14. It has been modified to work with R3.14. It includes a driver for an Ethernet GPIB interface. With this interface iocCore GPIB support is available on all supported R3.14 platforms.

## 4 SUPPORTING A NEW PLATFORM

It should be easy to port iocCore to additional environments as long as the environment supports multithreading, GNU make, and Perl.

Two parts of EPICS base must be extended: configure and libCom/osi.

EPICS base has a directory configure/os. For each supported platform, this directory contains files describing how to build components for that platform.

EPICS base has a directory src/libCom/osi/os that has the following structure:
- default
- posix
- \<platform specific\>

The osi directory contains definitions of the OSI interfaces. The implementation of these interfaces must be provided for each platform. The implementation can be provided in three ways.

- default – The directory default provides an implementation of many of the interfaces. If a default can be used, nothing needs to be done for the new platform.
- posix - The posix directory provides implementations for several of the interfaces but requires that the platform provide both POSIX real time and POSIX threads support. If the POSIX implementation of an interface can be used, then nothing needs to be done for the new platform.
- <platform specific> - If neither the POSIX nor default implementation will work for the new platform, then a platform-specific implementation must be provided.

Table 2 gives an idea of how much work is involved to support a new platform.

Table 2: Platform-Specific Lines of Code

| Implementation | Lines of Code |
|---|---|
| VxWorks | 1283 |
| RTEMS | 1488 |
| Solaris | 89 |
| Linux | 145 |
| Win32 | 3651 |
| Posix | 1262 |
| Default | 950 |

Lines of code include header, C, and C++ source files. The small number of lines of code required for Solaris and Linux is because both support POSIX real time and POSIX threads and both can use most of the default code. The Win32 implementation is much larger because the POSIX and most of the default code is not used.

## 5  OTHER 3.14 CHANGES

The Channel Access (CA) client interfaces are now thread safe on all operating systems (OSs), and the library now requires a multithreaded operating environment on all OSs. We hope that this uniformity will simplify testing and lead to better control over quality. The library now uses a preliminary plug-compatible interface when communicating with in-memory services such as the process control database. The code was restructured to not hold a mutual exclusion lock when calling user callbacks, thereby reducing subtle opportunities for deadlocks between in-memory services and the CA client library. Changes were also made to support unlimited vector lengths and client-specified dispatch priority in the server.

## 6  COMPATIBILITY

An important consideration in designing the iocCore port was compatibility for existing IOC applications and also for Channel Access clients.

Although EPICS base uses new build rules, the old rules are still supported, but only for vxWorks IOC applications. This allows easy conversion of existing applications to convert from R3.13 to R3.14.

Even though Channel Access was completely rewritten, the old CA client interface is still supported. Thus most existing CA clients are supported without any modifications.

## 7  CONCLUSION

The source code for EPICS iocCore was reorganized so that it can now be easily adapted to any operating environment with support for multi-threading. A new beta release is now available with these changes and other important upgrades including an improved build environment, a new command line shell, and communication of unlimited length arrays.

## REFERENCES

[1] M. Kraimer, "EPICS: Porting iocCore to Multiple Operating Systems," Proc. of the 1999 ICALEPCS Conference, Trieste, Italy, pp. 33-35, 2000.

[2] http://www.aps.anl.gov/epics/modules/base/R3-14.php